\documentstyle[floats,twocolumn,amssymb,aps,epsfig,pre]{revtex}
\draft

\begin{document}
\wideabs{ \title{Force Distributions in three dimensional compressible
granular packs} \author{J. Michael Erikson, Nathan W. Mueggenburg,
Heinrich M. Jaeger, Sidney R. Nagel} \address{The James Franck
Institute and Department of Physics\\ The University of Chicago \\
5640 S. Ellis Ave. Chicago, IL 60637} \date \today \maketitle

\begin{abstract}
We present an experimental investigation of the probability
distribution of normal contact forces, $P(F)$, at the bottom boundary
of static three dimensional packings of compressible granular
materials.  We find that the degree of deformation of individual
grains plays a large role in determining the form of this
distribution.  For small amounts of deformation we find a small peak
in $P(F)$ below the mean force with an exponential tail for forces
larger than the mean force.  As the degree of deformation is increased
the peak at the mean force grows in height and the slope of the
exponential tail increases.
\end{abstract}

}


It is known that forces within a granular material are distributed
in a highly inhomogeneous manner \cite{jaeger96}.  The largest
interparticle forces are arranged in a network of force chains while
other particles are shielded from the external force
\cite{dantu57,dantu67,howell97,ammi87,liu95}.  One quantitative way of
analyzing the inhomogenaities of these force networks is to measure
the probability distribution, $P(F)$, of normal forces, $F$, between
neighboring particles.

Experiments have shown that under a wide range of paramaters, $P(F)$
at the boundaries of granular packs decays exponentially for forces
larger than the mean force, $\overline F$, and has a small peak  near
the mean force (By ``small'' we mean that $P(F)$ increases by less
than a factor of two between its minimum near $F=0$ and the peak.)
\cite{liu95,mueth98,blair01,lovoll99,baxter97,makse00}. This form of
$P(F)$ has been found to be independent of interparticle friction and
the texture (geometrical ordering) of the granular pack.  Based upon
granular simulations and theoretical work it is expected that the form
of $P(F)$ should depend strongly on the amount of deformation of the
individual grains \cite{makse00,thornton97,thornton98,nguyen00}, with
a crossover to Gaussian behavior at high deformations.  Furthermore,
simulations of supercooled liquids (i.e., frictionless particles) by
O'Hern {\em et al.} suggest that the exponential tail in $P(F)$ might
arise from a self-averaging of configurations with different average
forces \cite{ohern01,ohern02}.  As the packing fraction is increased,
corresponding to greater deformations, they find that the relative
fluctuations in the average force decrease, leading to a Gaussian form
of $P(F)$.  Experimentally it has been difficult to measure the force
distribution of compressible materials.  In 2D shear experiments
Howell {\em et al.}  find a transition to Gaussian behavior for
deformations of the order of $2 \%$ \cite{howell99,howell-phone}.  In
3D static packings Makse {\em et al.} have reported some experimental
evidence for a transition from pure exponential to Gaussian.  However,
their maximum deformations were only of the order of $0.4 \%$
\cite{makse00} and conflict with the measurements of Blair {\em et
al.} at similar deformations \cite{blair01} and with the work of L\o
voll {\em et al.}  at very low amounts of deformation. \cite{lovoll99}.

We present an experimental investigation of the probability
distribution of normal forces at the boundaries of granular packs over
a wide range of deformations.  For average particle deformations up to
approximately $30 \%$ we find a form of $P(F)$ similar to that found
in previous experiments at low deformations as in references
\cite{mueth98,blair01,lovoll99}.  For large deformations (of the order
of $40 \%$) we find that $P(F)$ shows a much more pronounced peak
around the mean force.  Interestingly, we do not find Gaussian
behavior for large deformations.


Rubber beads of three different hardnesses ($40$, $50$, and $60$
durometer; hereafter referred to as soft, medium, and hard
respectively) with diameters $3.12 \pm 0.05$mm were contained within an
acrylic cylinder of inner diameter $140$mm.  Amorphous packings
approximately $72$mm in height were bounded on the top and bottom by
close fitting acrylic disks.  The packs of rubber beads were
constructed with one layer of glass beads at the bottom surface in a
crystalline arrangement.  Rubber beads were added on top of the glass
layer slowly so as not to disturb the underlying glass particles.  The
normal forces of the individual glass beads at the bottom surface were
measured using the carbon paper method
\cite{liu95,mueth98,blair01,delyon90}.  In this way the layer of glass
beads acted as an array of force transducers which could be easily
calibrated.

The experiments were performed by applying a force of between $2500$N
and $7000$N to the top piston of the cell with a hydraulic press.  The
normal forces between individual glass beads in the bottom layer and
the bottom piston were measured by placing carbon paper and white
paper \cite{materials} between the pack and the bottom piston.  The
size and intensity of the mark left on the white paper depended on the
magnitude of the normal force on the corresponding glass bead.

Following each experiment the white paper was carefully removed and
digitized with a flat bed scanner.  The images were then processed
using image analysis software to find the area and intensity of each
mark.  The intensities of the marks were converted to the force on the
corresponding bead using a fourth order polynomial interpolation of
calibration data as explained in reference \cite{blair01}.  An
appropriate number was added to the lowest bin to account for beads
with forces to small to leave a resolvable mark.  Since the bottom
layer was crystalline, the total number of contacts was known and
agreed well with the number of observed contacts at high applied
forces.  All forces for a given experimental run were normalized to
the average force for that run and the resulting probability
distribution, $P(f)$, of normalized forces, $f = F / \overline F$, was
averaged over $4$ to $11$ independent experimental runs.  Each run
provided approximately $1500$ imprints.  This results in a noise floor
of approximately $0.002$ to $0.003$ in $P(f)$ shown below.

The amount of deformations of individual grains was estimated by
measuring the deformation of individual grains in response to a known
compressional force.  The percent change in size along the direction
of the applied force was recorded as a function of applied force for
the three different hardnesses of rubber beads and for the glass
beads.  For the data shown below, the average degree of deformation is
listed.  This was obtained by measuring the compression of the
appropriate bead type in response to an applied force equal to
the average force of the corresponding distribution.

\begin{figure}[p]
\centerline{\epsfxsize= 8.0cm\epsfbox{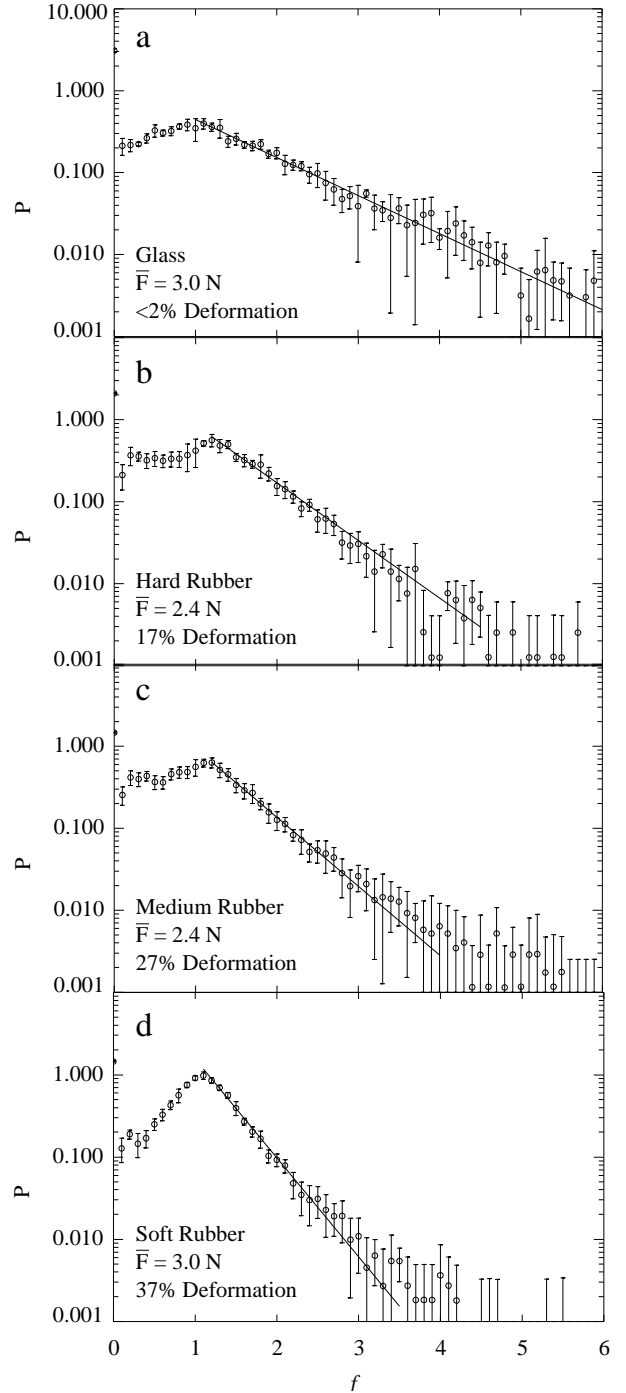}}
\vspace{.5cm}
\caption{$P(f)$ with varying particle hardness.  Probability
distribution of normal forces at the bottom boundary of amorphous
packing of a) glass, b) hard rubber, c) medium rubber, and d) soft
rubber beads.  Each plot represents an average of $4$ to $11$
experimental runs.  The average force per bead, $\overline F$, and
average deformation are indicated.  Error bars
represent statistical deviations from multiple experimental
realizations.  The solid line is a fit to an exponential over the
large force region resulting in slopes listed in table 1}
\label{dif_hardness}
\end{figure}

In order to compare our results with the earlier work of Mueth {\em et
al.} and Blair {\em et al.} we examined amorphous packings of smooth
spherical soda lime glass beads of diameter $3.06 \pm 0.04$mm.  Figure
\ref{dif_hardness}a shows the probability distribution of normal forces,
$P(f)$, at the bottom boundary averaged over four experimental runs
with an average force of $3.0$N per bead, corresponding to a
deformation of less than $2\%$.  We find an exponential decay
for large forces and a small peak in the distribution near the mean force,
consistent with previous experimental results \cite{mueth98,blair01,lovoll99}.

Experiments were next done with amorphous packs of rubber beads.
Figures \ref{dif_hardness}b,c, and d show the probability
distributions of normal forces at the bottom boundary for three
different hardnesses of beads with approximately the same average
force.  The hard and medium rubber packings show a distribution of
forces below the mean force similar to that of the glass beads, and
exhibit an exponential decay for forces larger than the mean force.
The shape of the distribution did not change significantly, however
the slopes of $-1.6$ and $-1.9$ respectively are somewhat larger than
the slope of $-1.1$ for the glass bead packings.  The soft rubber bead
packings, which had a larger amount of deformation, show a significantly
more pronounced peak and a steeper decay.  The region near the peak
remains exponential with a slope of $-2.8$, while above $f=2$ the
distribution departs from exponentail behavior.  Table \ref{table1}
shows the exponential decay constants for $P(f)$ at large forces and
the size of the peaks as characterized by the maximum height of the
peak divided by the minimum value of a smooth fit to the distribution
between $f = 0.1$ and $f = 0.7$.

\begin{table}[tb]
\begin{tabular}{lccccc}
Bead Type& $\overline F$ (N)&  Deformation ($\%$)& Slope&  Peak
Size\\
\hline \hline \\
Glass& $3.0$& $<2$& $-1.1$& $1.9$ \\
&&&&\\
Hard Rubber& $2.4$& $17$& $-1.6$& $1.8$ \\
&&&&\\
Medium Rubber& $2.4$& $27$& $-1.9$& $2.0$ \\
&&&&\\
Soft Rubber& $3.0$& $37$& $-2.8$& $6.0$ \\
&&&&\\
\hline 
&&&&\\
Soft Rubber& $1.6$& $25$& $-2.4$& $1.8$ \\
&&&&\\
Soft Rubber& $2.0$& $30$& $-2.6$& $2.0$ \\
&&&&\\
Soft Rubber& $3.0$& $37$& $-2.8$& $6.0$ \\
&&&&\\
Soft Rubber& $4.4$& $45$& $-3.8$& $29$ \\
\end{tabular}
\caption{\label{table1} Exponential decay constants for $P(f)$ at large
forces and Peak Size as explained in the text for various types of
beads at various levels of forcing and deformation.}
\end{table}

Note that changes in the calculated peak height are a good indicator
for changes in the overall shape of $P(f)$.  From Table 1 we see that
significant changes in peak height occur when the average deformation
exceeds roughly $30\%$.

\begin{figure}[p]
\centerline{\epsfxsize= 8.0cm\epsfbox{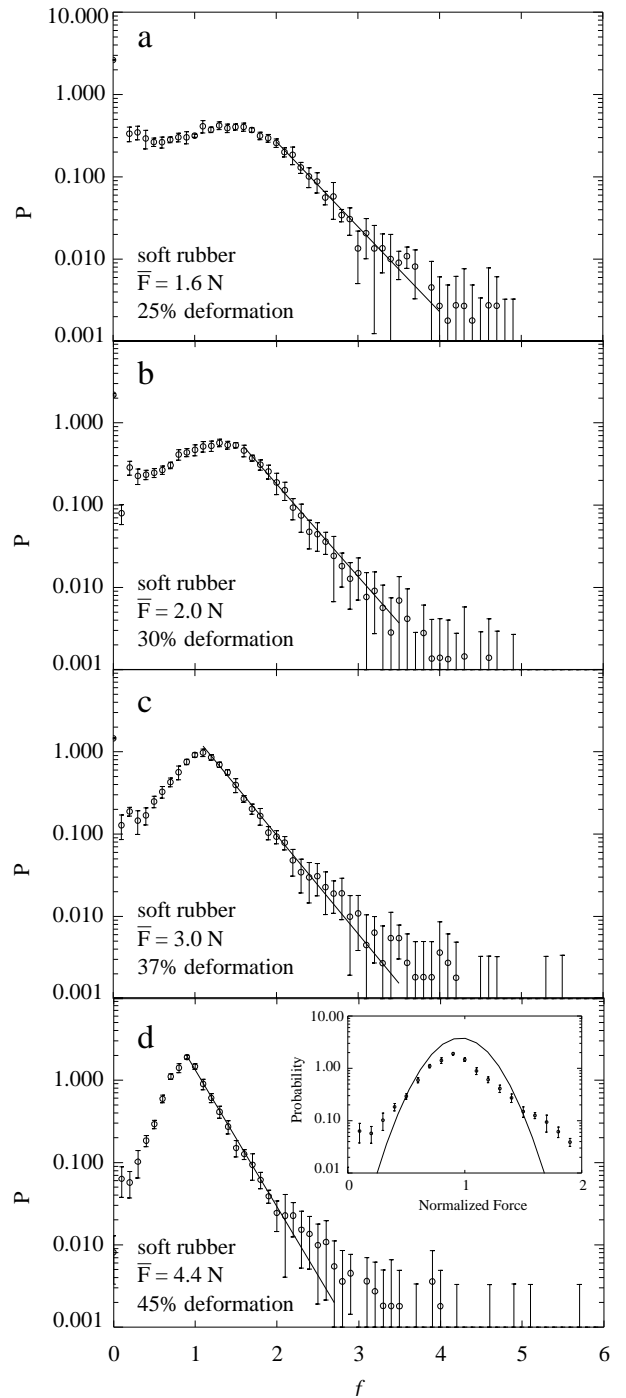}}
\vspace{.5cm}
\caption{$P(f)$ with varying applied force.  Probability distributions
of normal forces at the bottom boundary of amorphous packings of soft
rubber beads.  Each plot represents an average over $7$ to $9$
experimental runs.  Part c
is equivalent to figure \ref{dif_hardness}d. The error bars represent
statistical variations among experimental runs.  The solid lines are
fit to exponentials over the large force region.  The inset of d
compares the data to the Gaussian form obtained from a fit to $P(f)$
of a control experiment taken with a block of rubber on top of a
single layer of glass beads.}
\label{dif_press}
\end{figure}

To check this trend more directly, we performed the same experiments
for a single type of bead (soft rubber) with varying amounts of
pressure as shown in figure \ref{dif_press}.  As before, the slope of
$P(f)$ remains essentially unchanged (the peak height does not exceed
a value of 2) until the average degree of deformation exceeds roughly
$30\%$.  Beyond this amount of deformation, the peak size increases
sharply and $P(f)$ envolves into a much more symmetric form (Figure
\ref{dif_press}c, d).  Remarkably this evolution in the shape of
$P(f)$ does not seem to be connected with a change to Gaussian
behavior.  Near the peak, the distribution is well fit to an
exponential decay (see fitted lines in figures \ref{dif_hardness} and
\ref{dif_press}); at larger forces ($f>2$) the decay is actually
slower than exponential and shows the opposite trend to what would be
expected if the distribution was to revert to a Gaussian profile at
large deformations.

An intriging question is to what extent the force distribution of the
highly compressed rubber packings resembles that of a homogeneous
block of rubber.  The inset of figure \ref{dif_press}d compares data
from the main panel to a fit to data from a control experiment
performed on a block of rubber on top of a single layer of glass
beads.  The width of this fitted distribution is due to the resolution
of the carbon paper technique.  Note that at large deformations the
force distribution from a pack of rubber beads does not resemble that
of a rubber block, and in particular the distribution of forces from a
pack of rubber beads is significantly broader than that of a rubber
block.  Because of the limitations of the carbon paper technique we
are unable to rule out any residual influence of the single layer of
glass beads on the final shape of $P(f)$.  However, regardless of the
effect on the exact form of the probability distribution, any observed
changes in this distribution as the type of rubber bead is changed or
as the amount of deformation is increased must be connected to
properties of the rubber packing itself.


We find that the degree of deformation of individual particles does
play a large role in determining the form of the probability
distribution of forces within a granular pack.  When the degree of
deformation is small, either with hard particles or with soft
particles under a small force, we find that $P(f)$ has an exponential
decay for forces larger than the mean force and a small peak near the
mean force, consistent with previous experimental investigations
\cite{liu95,mueth98,blair01,lovoll99,baxter97,makse00}.  As the amount
of deformation increases beyond approximately $30\%$ the peak near the
mean force grows more pronounced.  This peaking behavior is in
agreement with simulations, although at higher deformations than would
have beeen expected
\cite{makse00,thornton97,thornton98,nguyen00,ohern01}.  For forces
larger than the mean force, we do not see a Gaussian decay.  The
distribution continues to decay exponentially (or possibly even
slower) at large forces.  If the single layer of glass beads at the
bottom surface does not alter the shape of the distribution then this
dependence is contrary to available simulation results and is as yet
unexplained.

We thank A. Bushmaker, E. Corwin, A. Marshall, M. M${\mathrm \ddot
o}$bius, and D. Mueth for their assistance with this project.  This
work was supported by NSF under Grant No. CTS-9710991, by the MRSEC Program
of the NSF under Grant No. DMR-9808595, and by the MRSEC REU program
at The University of Chicago.

\end{document}